\documentstyle[12pt]{article}

\setbox16= \hbox{der \kern -17.2pt \raise %
 3.8pt \hbox{-} \vrule width -2.9pt depth 0pt }
\def\dj{\copy16}

\begin{document}

\hoffset = -1truecm
\voffset = -2truecm

\thispagestyle{empty}

\begin{flushright}
LBL-33668 \\
UCB-PTH-93/10 \\
March 1993 \\
\end{flushright}
\vspace{1.0in}
\begin{center}
{\Large \bf
On the generality of certain predictions\\
for quark mixing \( \!\! ^{*} \)\\}
\vspace{1.0in}
{\bf Lawrence J. Hall \\}
\vspace{0.1in}
{\bf Andrija Ra\v{s}in \( \!\! ^{(a)} \) \\}
\vspace{0.1in}
Department of Physics, University of California, Berkeley, \\
and Lawrence Berkeley Laboratory, 1 Cyclotron Road, Berkeley, \\
California 94720,{\bf U.S.A.}
\end{center}

\baselineskip = 27pt

\begin{abstract}
The relations
${ {|V_{ub}|} \over {|V_{cb}|} } = \sqrt{ { m_u \over m_c } }$ and
${ {|V_{td}|} \over {|V_{ts}|} } = \sqrt{ { m_d \over m_s } }$
are significant successes of some specific models for quark masses.
We show that these relations are more general, resulting from
a much wider class of models. Consequences of these predictions
for CP violating asymmetries in neutral B meson decays
are discussed.

\vspace{25mm}

---------------------------------------------------------

\( \!\! ^{*} \)
This work was supported in part by the Director, Office of
Energy Research, Office of High Energy and Nuclear Physics, Division of
High Energy Physics of the U.S. Department of Energy under Contract
DE-AC03-76SF00098 and in part by the National Science Foundation under
grant PHY90-21139.

\end{abstract}

\newpage
\renewcommand{\thepage}{\roman{page}}

\begin{center}
{\bf Disclaimer}
\end{center}

\vskip .2in

\begin{scriptsize}
\begin{quotation}
This document was prepared as an account of work sponsored by the United
States Government.  Neither the United States Government nor any agency
thereof, nor The Regents of the University of California, nor any of their
employees, makes any warranty, express or implied, or assumes any legal
liability or responsibility for the accuracy, completeness, or usefulness
of any information, apparatus, product, or process disclosed, or represents
that its use would not infringe privately owned rights.  Reference herein
to any specific commercial products process, or service by its trade name,
trademark, manufacturer, or otherwise, does not necessarily constitute or
imply its endorsement, recommendation, or favoring by the United States
Government or any agency thereof, or The Regents of the University of
California.  The views and opinions of authors expressed herein do not
necessarily state or reflect those of the United States Government or any
agency thereof of The Regents of the University of California and shall
not be used for advertising or product endorsement purposes.
\end{quotation}
\end{scriptsize}

\vskip 2in

\begin{center}
\begin{small}
{\it Lawrence Berkeley Laboratory is an equal opportunity employer.}
\end{small}
\end{center}

\newpage
\renewcommand{\thepage}{\arabic{page}}
\setcounter{page}{1}


{\bf Motivation}
\vspace{7mm}

A theory of fermion masses should explain both the values of the quark and
lepton masses and the sizes of the four independent parameters of the
Kobayashi-Maskawa (KM) mixing matrix [1].
In the standard model these
quantities appear as free Yukawa
coupling parameters and must be determined
from experiment. While we are far
from a fundamental understanding of fermion
masses, theories which go beyond the standard model can possess symmetries
which reduce the number of free parameters of
these Yukawa coupling matrices, giving relationships between the
KM matrix elements and the quark masses. The first relationship so
obtained in a gauge theory was the very successful prediction for the Cabibbo
angle:
$|V_{us}| = \sqrt{ { m_d \over m_s } }$  [2], where $|V_{us}| = 0.221 \pm
0.002$
and $\sqrt{ { m_d \over m_s } } = 0.226 \pm 0.009$ [3]. Much interest has also
centered around the relation $|V_{cb}| = \sqrt{ { m_c \over m_t } }$ obtained
by Harvey, Ramond and Reiss [4] working with the form for the Yukawa
matrices written down by Georgi and Jarlskog [5]. If this relation
were valid at the weak scale the top quark would be predicted to be too heavy
[6]. However, inclusion of renormalization group (RG) corrections show that
such
a relation in a supersymmetric grand unified theory leads to a prediction of $
130 < m_t < 195$ GeV [7,8,9].

We choose $|V_{us}|$, $|V_{cb}|$,
${|V_{ub}| \over |V_{cb}|}$ and ${ |V_{td}| \over |V_{ts}|}$ as the
the four independent parameters of the KM matrix. Of these
$|V_{us}|$ and $|V_{cb}|$
are the two which are best measured. In this letter we concentrate on
predictions for
${|V_{ub}| \over |V_{cb}|}$ and ${ |V_{td}| \over |V_{ts}|}$. These are
predicted in several
schemes for fermion masses in terms of ratios of quark masses[10,11,6,7]
\begin{equation}
{ |V_{ub}| \over |V_{cb}| } = \sqrt{ { m_u \over m_c } } \simeq
0.061 \pm 0.009,
\end{equation}
and
\begin{equation}
{ |V_{td}| \over |V_{ts}| } = \sqrt{ { m_d \over m_s } } \simeq
0.226 \pm 0.009,
\end{equation}
where mass values from reference [3] have been used, keeping in mind
that the values in the ratios must be taken at the same renormalization
scale $\mu$.
In this letter we make two
comments about these relations: they are very successful, and they are quite
generic, following from a simple pattern for the Yukawa matrices.

The success of these relations has been magnified by last years announcement by
the CLEO collaboration [12] of lower values for ${|V_{ub}| \over |V_{cb}|}$.
They find central values of
${|V_{ub}| \over |V_{cb}| }$ of $0.053$, $0.062$, $0.065$ and $0.095$
in four phenomenological models used to analyze the data. The experimental
uncertainty is about $\pm0.020$.
Also the value of
the top quark mass obtained from precision electroweak data from LEP [13]:
$m_t = 145 \pm 25$ GeV is relevant because $|V_{td}|$ is probed
experimentally via the $B^0 - \bar{B}^0$ mixing parameter $x_d$ which is
strongly dependent on $m_t$:
\begin{eqnarray}
x_d & = &\tau_b { G_F^2 \over {6\pi^2} } ( \sqrt{B_B} f_B )^2  m_B \eta_B
m_t^2 S(y_t) Re(V_{td}^*V_{tb})^2
\nonumber \\
& = & 0.69
( { {\sqrt{B} f_B} \over {0.17GeV} } )^2
( {\eta_B \over 0.85} )
( {m_t \over {145 GeV} } )^2
( { {S(y_t)} \over 0.59 } )
( {{|V_{td}| \over |V_{ts}| } \over 0.226 } )^2
( {|V_{cb}| \over 0.043 } )^2,
\end{eqnarray}
where $y_t = m_t^2/M_W^2$,
$ S(y_t)=1-{3 \over 4}{ {y_t(1+y_t)} \over {(1-y_t)^2} } [ 1 + { {2y_t} \over
{1-y_t^2} } ln(y_t) ] $ and $\eta_B$ is the QCD correction factor.
{}From this it can be seen that by using central values for $m_t$ and other
parameters, together with the experimental result that $x_d = 0.70 \pm 0.10$,
the prediction of equation (2) is highly successful.

Given the success of these two predictions,
it is interesting to ask whether
they result from just a few specific models,
or whether they are generic
features of a wide class of theories [14].
In the rest of this letter we show
that predictions (1) and (2) occur whenever
two conditions on the elements of
the Yukawa matrices are satisfied.
We also show that CP violation measurements with neutral B mesons
will provide a test of whether the relations (1) and (2) provide a correct
understanding of $|V_{ub}|$ and $|V_{td}|$.

\vspace{7mm}

{\bf General constraint on the Yukawa matrices}
\vspace{7mm}

What conditions must the Yukawa matrices {\bf Y} ({\bf Y}={\bf U} or
{\bf D}) satisfy in order to get relations (1) and (2)?
The observed hierarchy of quark masses and mixing angles leads us to the
assumption that
the entries in the Yukawa matrices have a hierarchical structure, with $Y_{33}$
being the largest.
We first take $Y_{ij}$ to be real and later consider how the analysis is
modified by CP violating phases. The matrices {\bf Y} can be diagonalized by
three successive rotations in the (2,3), (1,3) and
(1,2) sectors (denoted by $s_{23},s_{13}$ and $s_{12}$ ):

\begin{eqnarray}
\lefteqn{
\left(
\begin{array}{ccc}
{\widetilde {\widetilde Y}}_{11} & 0 & 0 \\
0 & {\widetilde Y}_{22} & 0 \\
0 & 0 & Y_{33} \\
\end{array}
\right) =
\left(
\begin{array}{ccc}
1 & -s^{Y}_{12} & 0 \\
s^{Y}_{12} & 1 & 0 \\
0 & 0 & 1 \\
\end{array}
\right)
\left(
\begin{array}{ccc}
1 & 0 & -s^{Y}_{13} \\
0 & 1 & 0 \\
s^{Y}_{13} & 0 & 1 \\
\end{array}
\right)
\left(
\begin{array}{ccc}
1 & 0 & 0 \\
0 & 1 & -s^{Y}_{23} \\
0 & s^{Y}_{23} & 1 \\
\end{array}
\right) \times
}
\nonumber \\
& &
\times
\left(
\begin{array}{ccc}
Y_{11} & Y_{12} & Y_{13} \\
Y_{21} & Y_{22} & Y_{23} \\
Y_{31} & Y_{32} & Y_{33} \\
\end{array}
\right)
\left(
\begin{array}{ccc}
1 & 0 & 0 \\
0 & 1 & s'^{Y}_{23} \\
0 & -s'^{Y}_{23} & 1 \\
\end{array}
\right)
\left(
\begin{array}{ccc}
1 & 0 & s'^{Y}_{13} \\
0 & 1 & 0 \\
-s'^{Y}_{13} & 0 & 1 \\
\end{array}
\right)
\left(
\begin{array}{ccc}
1 & s'^{Y}_{12} & 0 \\
-s'^{Y}_{12} & 1 & 0 \\
0 & 0 & 1 \\
\end{array}
\right) .
\end{eqnarray}
The small rotation angles are given to leading order by
\begin{equation}
s^{Y}_{23} \simeq {Y_{23} \over Y_{33} } + { {Y_{32}Y_{22}} \over Y^2_{33} },\;
s'^{Y}_{23} \simeq {Y_{32} \over Y_{33} } + { {Y_{23}Y_{22}} \over Y^2_{33} },
\end{equation}
\begin{equation}
s^{Y}_{13} \simeq { {{\widetilde Y}_{13}} \over Y_{33} } +
{ {{\widetilde Y}_{31}Y_{11}} \over Y^2_{33} },\;
s'^{Y}_{13} \simeq{ {{\widetilde Y}_{31}} \over Y_{33} } +
{ {{\widetilde Y}_{13}Y_{11}} \over Y^2_{33} },
\end{equation}
\begin{equation}
s^{Y}_{12} \simeq{ {{\widetilde Y}_{12}} \over {{\widetilde Y}_{22}} } +
{ {{\widetilde Y}_{21}{\widetilde Y}_{11}} \over {{\widetilde Y}^2_{22}} },\;
s'^{Y}_{12} \simeq { {{\widetilde Y}_{21}} \over {{\widetilde Y}_{22}} } +
{ {{\widetilde Y}_{12}{\widetilde Y}_{11}} \over {{\widetilde Y}^2_{22}} }.
\end{equation}
The successive
rotations produce elements
\begin{equation}
{\widetilde {\widetilde Y}}_{11} \simeq
{\widetilde Y}_{11} -
{ {{\widetilde Y}_{12}{\widetilde Y}_{21}} \over {{\widetilde Y}_{22}} },\;
{\widetilde Y}_{11}  \simeq Y_{11} -
{ {{\widetilde Y}_{13}{\widetilde Y}_{31}} \over Y_{33} } ,\;
{\widetilde Y}_{22} \simeq Y_{22} - { {Y_{23}Y_{32}} \over Y_{33} },
\end{equation}
and
\begin{equation}
{\widetilde Y}_{12} = Y_{12} - Y_{13} s'^{Y}_{23},\;
{\widetilde Y}_{21} = Y_{21} - Y_{31} s^{Y}_{23},
\end{equation}
\begin{equation}
{\widetilde Y}_{13} = Y_{13} + Y_{12} s'^{Y}_{23},\;
{\widetilde Y}_{31} = Y_{31} + Y_{21} s^{Y}_{23}.
\end{equation}

The KM matrix which results from these rotations is
\begin{equation}
{\bf V} =
\left(
\begin{array}{ccc}
1  & s_{12} + s^{U}_{13}s_{23}  & s_{13} - s^{U}_{12}s_{23} \\
- s_{12} - s^{D}_{13}s_{23} & 1  & s_{23} + s^{U}_{12}s_{13} \\
- s_{13} + s^{D}_{12}s_{23}& - s_{23} - s^{D}_{12}s_{13} & 1 \\
\end{array}
\right),
\end{equation}
where $s_{23}=s^{D}_{23}-s^{U}_{23}$, $s_{13}=s^{D}_{13}-s^{U}_{13}$
and $s_{12}=s^{D}_{12}-s^{U}_{12}$.


To get relations (1) and (2)
it is sufficient to have:

$\bullet$ ${ |V_{ub}| \over |V_{cb}| } = |s^{U}_{12}|$ and
${ |V_{td}| \over |V_{ts}| } = |s^{D}_{12}|$ which is obtained by:
\begin{equation}
|s_{13}| << |s^{U}_{12}s_{23}| \,\, {\rm and} \,\, |s_{13}| <<
|s^{D}_{12}s_{23}|.
\end{equation}

$\bullet$ $|s^{U}_{12}| = \sqrt{ m_u \over m_c }$ and
$|s^{D}_{12}| = \sqrt{ m_d \over m_s }$ which is obtained by:

\begin{equation}
| {\widetilde Y}_{11}| <<
|{ {{\widetilde Y}_{12}{\widetilde Y}_{21}} \over {{\widetilde Y}_{22}} }|
\,\, {\rm and} \,\,
|{\widetilde Y}_{12}| = |{\widetilde Y}_{21}|.
\end{equation}
The conditions (12) and (13) on the Yukawa matrices
{\bf U} and {\bf D} allow for a wide class of mass ansatzes.
It is possible
that there are some other cases which will lead to (1) and (2) but we
believe that they would be quite special, involving for example nontrivial
cancellations.
All proposed ansatzes that we know of which lead to (1) and (2) satisfy
the conditions (12) and (13).

Now consider redoing the analysis with $Y_{ij}$ complex.The sequence of
rotations in (4) will now be interspersed with various diagonal rephasing
matrices. This will change the above equations in several ways. For example, in
(5), (6), (7) and (8)
the right hand sides of the
equations must be replaced by their
absolute values.
In equations (9) and (10) there
will be relative phases between the terms on the
right-hand sides. Finally the phase rotations will affect the KM matrix. While
the phase transformations cannot induce any new terms in $V_{ij}$, they can
multiply any of the existing ones by phases. However, it is clear that even in
this case equations (12) and (13) {\em are} the correct conditions for yielding
the predictions (1) and (2).

The conditions (12) and (13) are very simple, however, when expressed in terms
of $U_{ij}$ and $D_{ij}$ via equations (5) - (10), they appear quite
cumbersome.
Nevertheless, a simple heuristic way of stating the conditions on $U_{ij}$
and $D_{ij}$ is as follows:
\begin{itemize}
\item
$Y_{11}, Y_{13}$ and $Y_{31}$ must be small.

While conditions (12) and (13) are the precise statement on the smallness of
these elements, a feel for their meaning can be grasped as follows. The
smallness and hierarchy of fermion masses and mixings can
be restated in terms of approximate chiral and flavor symmetries which act on
each fermion type [15]. From these approximate
symmetries alone one finds that
the inequalities of (12) and (13)
become approximate equalities, thus
$|s_{13}| \approx |s^{U}_{12}s_{23}|$ etc. Hence these approximate chiral and
flavor symmetries are not sufficient to guarantee results (1) and (2). These
results follow only if the 11, 13 and 31 entries of the Yukawa matrices are
constrained by some more powerful means, for example by some new exact
symmetry. In many specific ansatzes, family symmetries force these to
vanish [10,11,6,7].
\item
$|{\widetilde Y}_{12}|  =  |{\widetilde Y}_{21}|$ usually results, to
sufficient
accuracy, whenever $| Y_{12}|  =  | Y_{21}|$.

While the 12 entries must be symmetric, other entries need not have
any symmetry.
None of the models
discussed in reference [16] has symmetric Yukawa matrices, but they all have
predictions (1) and (2).
\end{itemize}

The general conditions (12) and (13) are
satisfied by many special forms for
the Yukawa matrices, so that it is not
possible to use them to derive a
definite hierarchical structure for {\bf U} and {\bf D}.
However, this can be done for the subclass of
theories in which the Yukawa matrices
are also symmetric $Y_{ij}=Y_{ji}$,
and the resulting hierarchical
patterns are given in the appendix.

\vspace{7mm}

{\bf The KM matrix}
\vspace{7mm}

We now study the KM matrix which results from Yukawa couplings which satisfy
the conditions (12) and (13). In particular, the 11, 13 and 31 entries are
found to be
sufficiently small that they give only negligable corrections to the
diagonalization of {\bf U} and {\bf D}
\begin{equation}
{\bf L_U^\dagger} {\bf U} {\bf R_U} = {\bf U_d}~,~
{\bf L_D^\dagger} {\bf D} {\bf R_D} = {\bf D_d}.
\end{equation}
by the unitary matrices
\begin{equation}
{\bf L_U^\dagger} =
\left(
\begin{array}{ccc}
1 & -s_2 & 0 \\
s_2 & 1 & 0 \\
0 & 0 & 1 \\
\end{array}
\right)
\left(
\begin{array}{ccc}
e^{i\phi_U} & 0 & 0 \\
0 & 1 & 0 \\
0 & 0 & 1 \\
\end{array}
\right)
\left(
\begin{array}{ccc}
1 & 0 & 0 \\
0 & 1 & -s_U \\
0 & s_U & 1 \\
\end{array}
\right),
\end{equation}
\begin{equation}
{\bf L_D^\dagger} =
\left(
\begin{array}{ccc}
1 & -s_1 & 0 \\
s_1 & 1 & 0 \\
0 & 0 & 1 \\
\end{array}
\right)
\left(
\begin{array}{ccc}
e^{i\phi_D} & 0 & 0 \\
0 & 1 & 0 \\
0 & 0 & 1 \\
\end{array}
\right)
\left(
\begin{array}{ccc}
1 & 0 & 0 \\
0 & 1 & -s_D \\
0 & s_D & 1 \\
\end{array}
\right)
\left(
\begin{array}{ccc}
1 & 0 & 0 \\
0 & e^{-i\phi} & 0 \\
0 & 0 & 1 \\
\end{array}
\right),
\end{equation}
where $s_U$ and $s_D$ are $s_{23}$ rotations in the {\bf U} and {\bf D},
respectively, and $\phi_U, \phi_D$ and $\phi$ are the necessary
phase redefinitions.
This gives the KM matrix
\begin{equation}
{\bf V} =
{\bf L_U^\dagger L_D} =
\left(
\begin{array}{ccc}
e^{i\psi} & s_1e^{i\psi} - s_2e^{i\phi}  &
-s_2(s_De^{i\phi} - s_U) \\
s_2e^{i\psi} - s_1e^{i\phi} & e^{i\phi}  &
(s_De^{i\phi} - s_U) \\
-s_1(s_Ue^{i\phi} - s_D) & (s_Ue^{i\phi} - s_D) & 1 \\
\end{array}
\right),
\end{equation}
where $\psi = \phi_U - \phi_D$. For this matrix to yield the predictions (1)
and (2) we impose constraint (13), which implies
$|s_1| = \sqrt{m_d \over m_s}$ and $|s_2| = \sqrt{m_u \over m_c}$.
After an additional
phase redefinition, the KM matrix can be brought into the form
introduced in reference [7]:
\begin{equation}
{\bf V} =
\left(
\begin{array}{ccc}
1 & s_1 + s_2e^{-i\phi}  & s_2s_3 \\
-s_2 - s_1e^{-i\phi} & e^{-i\phi}  & s_3 \\
s_1s_3 & - s_3 & e^{i\phi} \\
\end{array}
\right),
\end{equation}
where $\phi\to\psi-\phi$ and $s_2\to-s_2$.
In (18) we do not loose any generality by choosing the corresponding angles
$\theta_1$, $\theta_2$ and $\theta_3$ to lie in the first quadrant.

This form is now seen to result directly from the
straightforward diagonalization of a large class of Yukawa textures.
It has the appealing feature that to a very good approximation $s_1, s_2$ and
$\phi$ are renormalization group invariants, while $s_3$ obeys a simple scaling
law. Note that while $s_1 = \sqrt{m_d \over m_s}$ and $s_2 =
\sqrt{m_u \over m_c}$ are given by quark masses and directly yield the
predictions (1) and (2), $s_3$ and $\phi$ are determined from $|V_{cb}|$ and
$|V_{us}|$, giving $s_3 = 0.043 \pm 0.007$ and, using the numbers quoted in (1)
and (2), $\sin \phi = 0.98{+0.02 \atop -0.07}$[17]. Hence we find that if
the Yukawa matrices satisfy (12) and (13), the entire KM matrix can be
determined quite accurately.

Since all four independent parameters of this KM matrix have now been specified
by CP conserving magnitudes $|V_{ij}|$, a crucial question is whether
the resulting prediction for CP violation agrees with data. Since $V_{us}$ in
(18) is not real, we use a rephase invariant result for the CP violating kaon
parameter $\epsilon$ [18]:
\begin{equation}
|\epsilon|= { G_F^2 \over {12\pi^2} } ( \sqrt{B_K} f_K )^2
{ m_K \over { \sqrt{2} \Delta m_K } } m_t^2
{ 1 \over { |\xi_u|^2 } }
\sum_{ij} S(y_i,y_j) Im(\xi_i \xi_j ( \xi_u^* )^2 ) \eta_{ij},
\end{equation}
where $y_i = m^2_i / m^2_W$, $\xi_i = V_{is}V_{id}^*$,
\begin{equation}
S(y_i,y_j) = { {y_iy_j} \over y_t }
\{ [ { 1 \over 4} + { 3 \over 2 } (1-y_j)^{-1} -
{ 3 \over 4 } (1-y_j)^{-2} ] { {lny_j} \over {y_j-y_i} }
+ ( y_j \to y_i ) - { 3 \over 4 } [ (1-y_i)(1-y_j) ]^{-1} \} ,
\end{equation}
and $\eta_{tt}=0.63$, $\eta_{ct}=0.34$ are the QCD correction factors.
Note that $S(y_t)=S(y_t,y_t)$.
Using central values we find
\begin{eqnarray}
|\epsilon| &=& 2.26 \cdot 10^{-3} \sin \phi
( { {\sqrt{B_K} f_K} \over {0.16GeV} } )^2
( { m_t \over 145GeV } )^2
( { 0.221 \over { |\xi_u| } } )^2 \times
\nonumber \\
&[&
 { { S(y_t) } \over 0.59 }
( { s_1 \over 0.226 } )^3
{ s_2 \over 0.061 }
( { s_3 \over 0.043 } )^4
 { {\eta_{tt}} \over 0.63 }
+ 0.12
 { { S(y_c,y_t) } \over {0.24 \cdot 10^{-3}} }
( { s_1 \over 0.226 } )^3
{ s_2 \over 0.061 }
( { s_3 \over 0.043 } )^2
{ {\eta_{ct}} \over 0.34 } ].
\end{eqnarray}
Here we used
$Im(\xi_t^2 (\xi_u^*)^2) \approx 2 s_1^3 s_2 s_3^4 sin\phi$ and
$Im(\xi_c \xi_t (\xi_u^*)^2) \approx s_1^3 s_2 s_3^2 sin\phi$.
We see that the experimental results indicate
a large (CP violating) phase $\phi$, consistent with its determination from
$|V_{us}|$.
An alternative way of stating the prediction for CP violation is via the
quantity J [19]:
\begin{equation}
J = Im ( V_{ud}V^*_{ub}V^*_{td}V_{tb} )
 = s_1s_2s^2_3sin\phi = \sqrt{m_d \over m_s}
\sqrt{m_u \over m_c} V^2_{cb} sin\phi = (2.6 \pm 0.9)10^{-5}sin\phi.
\end{equation}

\vspace{7mm}

{\bf CP asymmetries in B decays}
\vspace{7mm}

A good test of this KM matrix
comes from looking at the allowed
values for the CP asymmetries in
B decays [20]. The asymmetries,
given by  $\sin2\alpha$
(coming from $B_d \to \pi^+ \pi^-$) and
$\sin2\beta$ (coming from $B_d \to \psi K_S $),
can be expressed in terms of the Cabibbo angle $s_c \equiv |V_{us}|$,
$s_1$ and $s_2$[21]
\begin{equation}
\sin2 \alpha = - 2 \cos \phi \sin \phi,
\end{equation}
\begin{equation}
\sin2 \beta = { {2s_1 s_2 \sin \phi} \over {s_c^2} }
( 1 + { {s_2\cos\phi} \over s_1 } ).
\end{equation}

In the figure we plot the allowed region for $\sin2 \alpha$ and $\sin2 \beta$.
The dotted region is the region allowed by the standard model [22].
The $\sin2 \alpha$ variation comes mainly from the uncertainty in $s_1$ (i.e.
from the uncertainties in d and s masses), while the
$\sin2 \beta$ variation comes mainly from the uncertainty in $s_2$ (i.e.
from the uncertainties in u and c masses)[3].
Precise measurements of $\sin2\alpha$
and $\sin2\beta$ will reduce the experimental uncertainties on $s_1$ and
$s_2$, thereby providing a stringent test of (1) and (2).

We have shown that ${ |V_{ub}| \over |V_{cb}|}  = \sqrt{ { m_u \over m_c } }$
and
${ |V_{td}| \over |V_{ts}|}  = \sqrt{ { m_d \over m_s } }$ are highly
successful
relations which result from a wide range of models: the Yukawa matrices {\bf
U} and {\bf D} need only satisfy
the constraints (12) and (13). This typically means
that these matrices have small 11, 13, and 31 entries, and symmetric 12
entries.
Given the generality of these results, one might question whether
$ |V_{ub} / V_{cb}|$ and  $ |V_{td} / V_{ts}|$ can be used as a
probe of specific mass matrix ansatzes in future B-physics
experiments.
The answer is that they can, but
only if these schemes are able to predict $m_d/m_s$ and $m_u/m_c$ more
accurately than they are currently extracted from experiment.

\vspace{7mm}

{\bf Acknowledgement}

\vspace{7mm}

One of us (A.R.) would like to thank Uri Sarid for help
in preparation of the figure.
L.J.H. acknowledges partial support from the NSF Presidential Young
Investigator Program.

\newpage

{\bf Appendix. Forms of U and D with the additional constraint
$|Y_{ij}| = |Y_{ji}|$}
\vspace{7mm}

Let us try to find the most general
forms of {\it symmetric} {\bf U} and {\bf D}
which lead to (1) and (2). We neglect phases for simplicity.
As mentioned before, we assume no accidental cancelations,
so if a sum of two elements is small it is because they are
both small.
First, because of simmetricity, expressions (5)-(10) simplify:
$s^Y_{23}=s'^Y_{23}={Y_{23} \over Y_{33}}$ and therefore
\begin{equation}
{\widetilde Y}_{12}={\widetilde Y}_{21}=Y_{12}-Y_{13}{Y_{23} \over Y_{33}},
\end{equation}
\begin{equation}
{\widetilde Y}_{13}={\widetilde Y}_{31}=Y_{13}+Y_{12}{Y_{23} \over Y_{33}},
\end{equation}
\begin{equation}
s^Y_{13}=s'^Y_{13}={{\widetilde Y}_{13} \over Y_{33}},
\end{equation}
\begin{equation}
s^Y_{12}=s'^Y_{12}={{\widetilde Y}_{12} \over Y_{33}},
\end{equation}
and
\begin{equation}
{\widetilde {\widetilde Y}}_{11} \simeq
{\widetilde Y}_{11} -
{ {{\widetilde Y}^2_{12}} \over {{\widetilde Y}_{22}} },\;
{\widetilde Y}_{11}  \simeq Y_{11} -
{ {{\widetilde Y}^2_{13}} \over Y_{33} } ,\;
{\widetilde Y}_{22} \simeq Y_{22} - { {Y^2_{23}} \over Y_{33} }.
\end{equation}
We will express all mass matrix
elements in terms of their eigenvalues
(recall that $m_1={\widetilde {\widetilde Y}}_{11}$,
$m_2={\widetilde Y_{22}}$ and $m_3=Y_{33}$).
Because of our assumption of no accidental cancellations
we divide possible forms into two categories: either
$Y_{22}<<m_2$ and $Y_{23}=\sqrt{m_2m_3}$,
or
$Y_{22}=m_2$ and $Y_{23}<<\sqrt{m_2m_3}$.

Now we use conditions (12) and (13).
Let us first use condition (13) because it does not depend on whether
{\bf Y} is {\bf U} or {\bf D}. It tells us that
\begin{equation}
{\widetilde Y_{12}}=\sqrt{m_1m_2},
\end{equation}
\begin{equation}
Y_{11} << m_1,
\end{equation}
\begin{equation}
{\widetilde Y_{13}} << \sqrt{m_1m_3}.
\end{equation}
Since $Y_{23} \leq \sqrt{m_2m_3}$
(from $m_2={\widetilde Y}_{22}$), it follows from equations
(30) and (32) that $Y_{12}=\sqrt{m_1m_2}$ and
$Y_{13}<<\sqrt{m_1m_3}$.
Therefore the
symmetric {\bf U} and {\bf D} that obey (13)
must take one of the following forms:
\begin{equation}
\left(
\begin{array}{ccc}
(Y_{11} << m_1)  & \sqrt{m_1m_2}  &  (Y_{13} << \sqrt{m_1m_3}) \\
\sqrt{m_1m_2} & (Y_{22}<< m_2)  & \sqrt{m_2m_3} \\
(Y_{13} << \sqrt{m_1m_3}) & \sqrt{m_2m_3} & m_3 \\
\end{array}
\right),
\end{equation}
where $Y_{22} << {Y_{23}^2 \over Y_{33}}$, or
\begin{equation}
\left(
\begin{array}{ccc}
(Y_{11} << m_1)  & \sqrt{m_1m_2}  &  (Y_{13} << \sqrt{m_1m_3}) \\
\sqrt{m_1m_2} & m_2  & (Y_{23} << \sqrt{m_2m_3}) \\
(Y_{13} << \sqrt{m_1m_3}) & (Y_{23} << \sqrt{m_2m_3}) & m_3 \\
\end{array}
\right),
\end{equation}
where $Y_{22} >> {Y_{23}^2 \over Y_{33}}$.

Let us now use the constraint (12)
which will further constrain some of the
bracketed elements in (33) or (34).
Using reasonable values for quark masses we see
that in (12) the more stringent constraint is
\begin{equation}
s_{13}<< \sqrt{u_1 \over u_2}s_{23},
\end{equation}
where $s_{13}=s^D_{13}-s^U_{13}$ and $s_{23}=s^D_{23}-s^U_{23}$.
We use $u_1=m_u$, $u_2=m_c$, etc.
In particular, both $s^U_{13}$ and $s^D_{13}$ must be less than
$\sqrt{u_1 \over u_2}s_{23}$:
\begin{equation}
|{U_{13} \over u_3} + {\sqrt{u_1u_2} \over u_3}{U_{23} \over u_3}|
<< \sqrt{u_1 \over u_2} | {D_{23} \over d_3} - {U_{23} \over u_3} |,
\end{equation}
\begin{equation}
|{D_{13} \over d_3} + {\sqrt{d_1d_2} \over d_3}{D_{23} \over d_3}|
<< \sqrt{u_1 \over u_2} | {D_{23} \over d_3} - {U_{23} \over u_3} |.
\end{equation}
Notice that consistency of solutions is automatically obeyed since
${\sqrt{d_1d_2} \over d_3} << \sqrt{u_1 \over u_2}$ and
${\sqrt{u_1u_2} \over u_3} << \sqrt{u_1 \over u_2}$ for reasonable
quark masses. Therefore, we conclude that limits on $U_{13}$ and
$D_{13}$ may be somewhat stringent
\begin{equation}
U_{13} << min\{ \sqrt{u_1u_3}, \sqrt{u_1 \over u_2}u_3s_{23} \} \equiv a\, ,
\end{equation}
\begin{equation}
D_{13} << min\{ \sqrt{d_1d_3}, \sqrt{u_1 \over u_2}d_3s_{23} \} \equiv b\, .
\end{equation}
If {\bf U} or {\bf D} is of type (34) somewhat stringent limits
on $U_{23}$ and $D_{23}$ are also possible
\begin{equation}
U_{23} << min\{ \sqrt{u_2u_3},
{ u^2_3 \over u_2} s_{23} \} \equiv c \, ,
\end{equation}
\begin{equation}
D_{23} << min\{ \sqrt{d_2d_3},
\sqrt{u_1 \over u_2}{ d^2_3 \over \sqrt{d_1d_2}} s_{23} \} \equiv d\, .
\end{equation}

We can now write possible forms of symmetric {\bf U} and {\bf D}
which lead to successful predictions (1) and (2). There are four
possibilities depending on whether {\bf U} or {\bf D}
take on the form (33) or (34)[23]

1)
\begin{equation}
{\bf U} =
\left(
\begin{array}{ccc}
(U_{11} << u_1)  & \sqrt{u_1u_2}  &
(U_{13} << \sqrt{u_1u_3}) \\
\sqrt{u_1u_2} & (U_{22}<< u_2)  & \sqrt{u_2u_3} \\
(U_{13} << \sqrt{u_1u_3}) & \sqrt{u_2u_3} & u_3 \\
\end{array}
\right),
\end{equation}
\begin{equation}
{\bf D} =
\left(
\begin{array}{ccc}
(D_{11} << d_1)  & \sqrt{d_1d_2}  &
(D_{13} << \sqrt{d_1d_3}
\sqrt{ { {u_1d_2} \over {u_2d_1} } })\\
\sqrt{d_1d_2} & (D_{22}<< d_2)  & \sqrt{d_2d_3} \\
(D_{13} << \sqrt{d_1d_3}
\sqrt{ { {u_1d_2} \over {u_2d_1} } })
& \sqrt{d_2d_3} & d_3 \\
\end{array}
\right).
\end{equation}

2)
\begin{equation}
{\bf U} =
\left(
\begin{array}{ccc}
(U_{11} << u_1)  & \sqrt{u_1u_2}  &
(U_{13} << \sqrt{u_1u_3}) \\
\sqrt{u_1u_2} & u_2  & (U_{23} << \sqrt{u_2u_3}) \\
(U_{13} << \sqrt{u_1u_3})
& (U_{23} << \sqrt{u_2u_3}) & u_3 \\
\end{array}
\right),
\end{equation}
\begin{equation}
{\bf D} =
\left(
\begin{array}{ccc}
(D_{11} << d_1)  & \sqrt{d_1d_2}  &
(D_{13} << \sqrt{d_1d_3}
\sqrt{ { {u_1d_2} \over {u_2d_1} } })\\
\sqrt{d_1d_2} & (D_{22}<< d_2)  & \sqrt{d_2d_3} \\
(D_{13} << \sqrt{d_1d_3}
\sqrt{ { {u_1d_2} \over {u_2d_1} } })
& \sqrt{d_2d_3} & d_3 \\
\end{array}
\right).
\end{equation}

3)
\begin{equation}
{\bf U} =
\left(
\begin{array}{ccc}
(U_{11} << u_1)  & \sqrt{u_1u_2}  &  (U_{13} << a)\\
\sqrt{u_1u_2} & (U_{22}<< u_2)  & \sqrt{u_2u_3} \\
(U_{13} << a)& \sqrt{u_2u_3} & u_3 \\
\end{array}
\right),
\end{equation}
\begin{equation}
{\bf D} =
\left(
\begin{array}{ccc}
(D_{11} << d_1)  & \sqrt{d_1d_2}  & (D_{13} << b) \\
\sqrt{d_1d_2} & d_2  & (D_{23} << d) \\
(D_{13} << b)& (D_{23} << d) & d_3 \\
\end{array}
\right).
\end{equation}

4)
\begin{equation}
{\bf U} =
\left(
\begin{array}{ccc}
(U_{11} << u_1)  & \sqrt{u_1u_2}  &  (U_{13} << a)\\
\sqrt{u_1u_2} &  u_2  & (U_{23} << c) \\
(U_{13} << a) & (U_{23} << c) & u_3 \\
\end{array}
\right),
\end{equation}
\begin{equation}
{\bf D} =
\left(
\begin{array}{ccc}
(D_{11} << d_1)  & \sqrt{d_1d_2}  & (D_{13} << b) \\
\sqrt{d_1d_2} & d_2  & (D_{23} << d) \\
(D_{13} << b)& (D_{23} << d) & d_3 \\
\end{array}
\right),
\end{equation}
where a,b,c and d are given in equations (38)-(41).

In the above it
is understood that $U_{23}$ and
$D_{23}$ cannot be simultaneously =0 since they are constrained by the
condition $V_{cb}=s_{23}$.

Some specific mass matrix ansatzes can be
recovered by setting bracketed elements to zero. For example, 1) contains
the Fritzsch scheme[10], while 2) is the generalization of the
Harvey,Reiss and Ramond form[4]. Nevertheless, it is important to notice
that although the bracketed elements can in many cases be set to zero they
need not to be. As long as they obey the limits, relations (1) and (2) will
follow. For example, $Y_{13}$ can be as big as $Y_{12}$!

\newpage

\( ^{(a)} \) On leave of absence from the
Ru\dj \hspace{.1in} Bo\v{s}kovi\'{c} Institute, Zagreb, Croatia.



\begin{thebibliography}{99}

\bibitem[1] MM. Kobayashi and T. Maskawa,
Prog. Theor. Phys. {\bf 49}, 652 (1973).
\bibitem[2] SS. Weinberg in {\it A Festschrift for I.I. Rabi},
edited by L. Motz (New York Academy of Sciences, New York 1977),
F. Wilczek and A.
Zee, Phys. Lett. {\bf 70B} 418 (1977), H. Fritzsch, Phys. Lett. {\bf 70B} 436
(1977). For a historical review of mass relations and further references see
L.J. Hall, preprint LBL-33161 (1992), to appear in the
Proceedings of the 1992 SLAC Summer School
Proceedings.
\bibitem[3] FFor definiteness, in this paper we use values
for the current quark masses of
$m_u(1 GeV) = 5.1 \pm 1.5$ MeV, $m_s/m_d = 19.6 \pm 1.6$, and $m_c(m_c) = 1.27
\pm 0.05$ GeV, taken from
 J. Gasser and H. Leutwyler, Phys. Rep. {\bf 87}, 82 (1982).
There has been
considerable debate as to whether these are indeed the correct values to use.
The results of this paper can be simply scaled for other values.
\bibitem[4] JJ.A. Harvey, D.B. Reiss
and P. Ramond, Phys. Lett. {\bf 92B}, 309 (1989);
Nucl. Phys. {\bf B199}, 223 (1982).
\bibitem[5] HH. Georgi and C. Jarlskog, Phys. Lett. {\bf 86B}, 297 (1979).
\bibitem[6] XX.G. He and W.S. Hou, Phys. Rev. {\bf D41} 1517 (1990).
\bibitem[7] SS. Dimopoulos, L. J. Hall and S. Raby, Phys. Rev. Lett. {\bf 68}
1984 (1992); Phys. Rev. {\bf D45}, 4192 (1992).
\bibitem[8] PP. Ramond, UFIFT-92-4 (1992); H. Arason, D.J. Casta\~{n}o,
E. J. Piard and P. Ramond, Phys. Rev. {\bf D47}, 232 (1993).
\bibitem[9] GG. Anderson, S. Dimopoulos, L.J. Hall and S. Raby, preprint
OHSTPY-HEP-92-018, (1992), to appear in Phys. Rev. {\bf D}; V. Barger,
M.S. Berger, T. Han and M. Zralek, Phys. Rev. Lett. {\bf 68}, 3394 (1992);
V. Barger, M.S. Berger and P. Ohmann, preprint MAD-PH-711 (1992).
\bibitem[10] HH. Fritzsch, Nucl. Phys. {\bf B155}, 189 (1979).
\bibitem[11] BB. Stech, Phys. Lett. {\bf 130B}, 189 (1983).
\bibitem[12] PP. S. Drell (CLEO collaboration), talk given at
the Dallas Conference (1992); D. Cassel, Review talk, DPF 92 Meeting, Fermilab
(Nov. 1992).
\bibitem[13] GG. Altarelli, R. Barbieri, S. Jadach, Nucl. Phys. {\bf B369},
3 (1992);
L. Rolandi (LEP), talk given at the Dallas Conference (1992).
\bibitem[14] WWe thank S. Dimopoulos
for asking this question.
\bibitem[15] CC. D. Froggatt and H. B. Nielsen, Nucl. Phys. {\bf B147}
277 (1979);
A. Antaramian, L. J. Hall and A. Ra\v{s}in, Phys. Rev. Lett.
{\bf 69}, 1871 (1992); L. J. Hall and S. Weinberg, preprint UTTG-22-92
(1993).
\bibitem[16] GG. Anderson, S. Dimopoulos, L. J. Hall, S. Raby and G. Starkman,
preprint LBL-33531 (1993).
\bibitem[17] NNote that the experimental data on
${\rm Re}\epsilon$ in the kaon system forces $sin\phi$ to be positive,
so that $\phi$ lies in the first or second quadrant.
\bibitem[18] SSee I. Dunietz, Ann. Phys. {\bf 184}, 380 (1988)
and references therein.
\bibitem[19] CC. Jarlskog, Phys. Rev. Lett. {\bf 55}, 1039 (1985).
\bibitem[20] AA. Carter and A. Sanda, Phys. Rev. Lett. {\bf 45}, 952 (1980);
Phys. Rev. {\bf D 23}, 1567 (1981);
For a recent review see Y. Nir
and H. Quinn, preprint SLAC-PUB-5737 (1992).
\bibitem[21] SS. Dimopoulos, L. J. Hall and
S. Raby, Phys. Rev. {\bf D46}, 4793 (1992).
\bibitem[22]YY. Nir and U. Sarid, preprint WIS-92-52-PH (1992), to appear
in Phys. Rev. {\bf D}.
\bibitem[23] NNotice that when {\bf D} is of type (33) then
$s_{23}=\sqrt{d_2 \over d_3}$ and when reasonable quark
masses are used, it is easy to see that the additional constraint (12)
affects only $D_{13}$ ($U_{13}$,$U_{23}$ and $D_{23}$ are more constrained
by (13)).
\end{thebibliography}
\end{document}